\documentclass[%
 reprint,
amsmath,amssymb,
aps,
]{revtex4-2}

\usepackage{bm}
\usepackage{siunitx}

\usepackage{mathrsfs}
\usepackage{eqnarray,amsmath}
\usepackage{graphicx}
\usepackage{amssymb}
\usepackage{textcomp}
\usepackage{color}
\usepackage{xcolor}
\usepackage{natbib}
\usepackage{dcolumn}
\usepackage{ragged2e}
\usepackage{array}
\usepackage{dcolumn}
\usepackage{appendix}
\usepackage{etoolbox}

\usepackage{xr}
\usepackage{hyperref}
\definecolor{txtNB}{RGB}{181,12,61}  
\definecolor{txtOD}{RGB}{235,0,31}  
\definecolor{txtAP}{RGB}{18,144,69} 

\begin{document}

\title{Ionization Energy of Rb$_2$ by electric field-ionization of molecular Rydberg states}


\author{Manuel Alejandro Lefrán Torres}
\affiliation{Instituto de F\'{i}sica de S\~{a}o Carlos, Universidade de S\~{a}o Paulo, Caixa Postal 369, 13560-970, S\~{a}o Carlos, SP, Brasil}

\author{David Rodríguez Fernández}
\affiliation{Instituto de F\'{i}sica de S\~{a}o Carlos, Universidade de S\~{a}o Paulo, Caixa Postal 369, 13560-970, S\~{a}o Carlos, SP, Brasil}

\author{Jaime Javier Borges Márquez}
\affiliation{Instituto de F\'{i}sica de S\~{a}o Carlos, Universidade de S\~{a}o Paulo, Caixa Postal 369, 13560-970, S\~{a}o Carlos, SP, Brasil}

\author{Marcos Roberto Cardoso}
\affiliation{Instituto de F\'{i}sica de S\~{a}o Carlos, Universidade de S\~{a}o Paulo, Caixa Postal 369, 13560-970, S\~{a}o Carlos, SP, Brasil}

\author{Luis Gustavo Marcassa}
\email{marcassa@ifsc.usp.br}
\affiliation{Instituto de F\'{i}sica de S\~{a}o Carlos, Universidade de S\~{a}o Paulo, Caixa Postal 369, 13560-970, S\~{a}o Carlos, SP, Brasil}

\author{Amrendra Pandey}
\affiliation{Université Paris-Saclay, CNRS, Laboratoire Aimé Cotton, 91400 Orsay, France}

\author{Romain Vexiau}
\affiliation{Université Paris-Saclay, CNRS, Laboratoire Aimé Cotton, 91400 Orsay, France}

\author{Olivier Dulieu}
\affiliation{Université Paris-Saclay, CNRS, Laboratoire Aimé Cotton, 91400 Orsay, France}

\author{Nadia Bouloufa-Maafa}
\affiliation{Université Paris-Saclay, CNRS, Laboratoire Aimé Cotton, 91400 Orsay, France}

\date{\today}%

\begin{abstract}
We report the measurement of the ionization energy of the $^{85}\text{Rb}_2$ molecule through resonantly enhanced 2-photon ionization in a supersonic beam. The first photon excites the $X^1\Sigma_g^+ (v_X = 0)\rightarrow B^1\Pi_u (v_B = 2)$ transition, while the second photon wavenumber is scanned over the 16720~cm$^{-1}$-16750~cm$^{-1}$ range, thus yielding a structured spectrum of Rb$_2^+$ ions extracted by an electric field and recorded by mass spectrometry. We modeled the onset of the ionization signal as a function of the electric field strength between 18~V/cm and 180~V/cm, leading to the Rb$_2$ ionization energy $E_i = 31497.3 \pm 0.6$~cm$^{-1}$, and to the dissociation energy of the Rb$_2^+$ ground state  $D_0 = 6158.2 \pm 0.6$ cm$^{-1}$. Our measured value $E_i$ is found to be $149.3$~cm$^{-1}$ larger than the one reported in the experiment by Bellos \textit{et al.} [Phys. Rev. A \textbf{87}, 012508 (2013)]. Our value of $D_0$ agrees with our theoretical determination using a quantum chemistry approach. Using a simple theoretical model, we assign unevenly spaced structures of the ionization spectrum to molecular Rydberg levels belonging to several series that converge to the lowest vibrational levels of Rb$_2^+$. 
\end{abstract}


\maketitle
\section{Introduction}

Nowadays, researches on ultracold ($T \ll 1\,\mu$K) and dilute ($n \approx 10^8-10^{13}$~cm$^{-3}$) quantum gases are routinely merging atoms, molecules, and ions to study various regimes of competing interactions between quantum particles \cite{karpa2025}. For such studies, alkali-metal atoms are the exquisite but not exclusive choice because of their simplicity and convenient optical properties, and among them, the rubidium atom. For example, two recent series of experiments explored the behavior of a Rb quantum gas exposed to the presence of a charged impurity Rb$^+$. The detected products are Rb$_2$ and Rb$_2^+$ molecules. A first series of experiments \cite{Harter2012,Schmid2010,Schmid2012} has indirectly observed the formation of deeply bound Rb$_2$ molecules by monitoring the decay of an atomic cloud when a single $^{138}$Ba$^+$ or $^{87}$Rb$^+$ ions are brought to interaction with ultracold thermal or Bose-condensed $^{87}$Rb gas. In a second series \cite{Dieterle2020,Dieterle2021}, weakly bound Rb$_2^{+}$ ions have been observed after a single cold $^{87}$Rb$^{+}$ ion was injected into a $^{87}$Rb Bose-Einstein condensate. In both series, it has been proven that three-body recombination dominates the dynamics of the gas \cite{pandey2025}. 

Modeling of such experiments requires precise knowledge of the $X^2\Sigma_g^+(v)$ ground state potential energy curve of Rb$_2^+$ and thus the ionization energy $E_i$ of Rb$_2$. Although numerous theoretical investigations have been reported (see \cite{schnabel2022high,pandey2024ultracold,dasilva2024} and references therein), the experimental value of $E_i$ has long been poorly known for Rb$_2$ (with an uncertainty of 400~cm$^{-1}$), in contrast to other alkali-metal dimers \cite{stwalley1993b} (with typical uncertainty lower than 3~cm$^{-1}$). In 2013, Bellos \textit{et al.} \cite{bellos2013upper} reported a precise upper bound for $E_i$ (with an uncertainty of 0.6~cm$^{-1}$), based on photoassociation spectroscopy of autoionizing electronic states of Rb$_2$.

In the present work, we measured the ionization energy of $^{85}$Rb$_2$ in a supersonic beam, using resonantly enhanced 2-photon ionization (RE2PI). We exposed ground-state molecules to a first laser to reach a well-known intermediate excited state, while the wavenumber of a second laser was scanned over a range where ionization is expected according to the studies above, either by direct photoionization or by pulsed-field ionization of molecular Rydberg levels. With this RE2PI spectroscopy, we extracted a value for $E_i$, and the corresponding binding energy $D_0(X^+)$ of the $^{85}$Rb$_2^+$ ground state ion, with an uncertainty of 0.6~cm$^{-1}$.

In Section \ref{sec:excitation} we recall the main aspects of our molecular structure calculations required for the modeling of the two-photon transition in Rb$_2$ and we describe in Section \ref{sec:experiment} the experimental setup. We present in Section \ref{sec:spectra} the Rb$_2^+$ spectra as functions of the wavenumber of the second laser, recorded for several electric field strengths, where we infer the presence of molecular $^{85}$Rb$_2$ Rydberg levels based on a simple model. In Section \ref{sec:ionization}, the $^{85}$Rb$_2$ ionization energy is extracted from the evaluation of the onset of ionization measured as a function of the applied external electric field strength. We conclude the paper with a discussion of the good agreement between the measurements and the theoretical calculations, while we propose an explanation for the large discrepancy observed with the value of \cite{bellos2013upper}.

\section{Excitation scheme}
\label{sec:excitation}

\begin{figure}[t]
\centering
\includegraphics[width=0.425\textwidth]{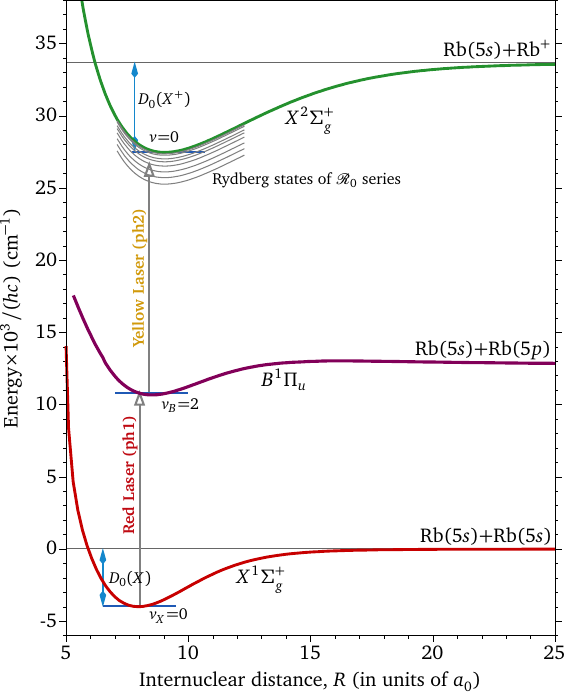}
\caption{\label{Fig_PECs} (Color online) Rb$_2$ potential energy curves relevant for the present experiment. The red photon (ph1) excites the $X^1\Sigma_g^+ (v_X = 0)\rightarrow B^1\Pi_u (v_B = 2)$ transition at 14753 cm$^{-1}$. The yellow photon (ph2) is scanned in the  $16720$~cm$^{-1}$-$16755$~cm$^{-1}$ wavenumber range, around the expected minimum of the $X^2\Sigma_g^+$ PEC. Grey lines illustrate the presence of Rb$_2$ Rydberg states, drawn as shifted  $X^2\Sigma_g^+$ PEC. }
\end{figure}

The two-photon excitation scheme (Fig. \ref{Fig_PECs}) is applied to Rb$_2$ molecules prepared at the lowest vibrational level $v_X=0$ of the electronic ground state $X^1\Sigma_g^+$ (hereafter referred to as the $X$ state). The first photon (at 14753~cm$^{-1}$) excites the Rb$_2$ molecules toward the vibrational level $v_B=2$ of the excited electronic state $B^1\Pi_u$ (hereafter referred to as the $B$ state). The second laser is scanned between 16720~cm$^{-1}$ and 16755~cm$^{-1}$ to locate the position of the ionization threshold of Rb$_2$. For our simulation of the ionization spectrum, we use a combination of PECs extracted from previous experimental work, and from our own computations. The PEC of the Rb$_2$ $X$ state is taken from the spectroscopic analysis of \cite{amiot1990laser,strauss2010hyperfine}. The PEC for the Rb$_2$ $B$ state is the result of a combination of spectroscopic data \cite{amiot1990laser} and our calculations \cite{beuc2007absorption}. Note that the transition between $v_X=0$ and $v_B=2$ is well known experimentally \cite{amiot1990laser} (see also the next Section). The PEC of the ground state $X^2\Sigma_g^+$ of Rb$_2^+$ (hereafter referred to as the $X^+$ state) is taken from our calculations \cite{aymar2003model}, which have been updated in \cite{beuc2007absorption}. The methodology is described in detail in \cite{aymar2005calculation, vexiau2017dynamic}. The Rb$_2^+$ and Rb$_2$ are described, respectively, as a one- and two-valence electron system, moving in the field of effective core potentials (ECP) completed with core polarization potentials to account for core valence correlation. The Rb$_2^+$ PECs are the result of a simple Hartree-Fock calculation, while a full configuration interaction (FCI) is achieved for the Rb$_2$ PECs. All ECP and CPP parameters, as well as the basis sets, are reported in the quoted references. By construction of our theoretical approach, the binding energy of the $5s$ level and the spin-averaged excitation energy of the $5p$ level of Rb are identical to the experimental ones \cite{johansson1961spectra,menegatti2013trap}. Taking the Rb($5s$)+Rb($5s$) asymptote as the origin of energy, the energy of the Rb($5s$)+Rb($5p$) and Rb($5s$)+Rb$^+$ dissociation limits are $hc \times 12737.343$~cm$^{-1}$, and $hc \times 33690.810$~cm$^{-1}$, respectively. The main spectroscopic properties of the $X$ and $B$ PECs of Rb$_2$, and of the $X^+$ PEC of Rb$_2^+$ are collected in Table \ref{tab:Rb2pmin}. 

\setlength{\tabcolsep}{1.0pt}
\begin{table}[h]
\begin{tabular}{ccccc}
\hline \hline
Ref.&$D_e$ & $\omega_e$ & $B_e$ & $D_0$\\
(cm$^{-1}$) &(cm$^{-1}$) &(cm$^{-1}$)&(cm$^{-1}$)&(cm$^{-1}$) \\
\hline \hline
\multicolumn{5}{c}{$X^1\Sigma_g^+$}\\
\cite{strauss2010hyperfine},\cite{amiot1990laser}&3993.5928(30)     &57.749(2)&0.0223973(4)&3964.718  \\
t$_{12}$ &3993.43  &57.50 &0.022&3964.68\\
t$_{11}$ &3993.88  &56.69 &0.022&3965.54\\ \hline
\multicolumn{5}{c}{$B^1\Pi_u$ }\\
\cite{amiot1990laser}    &   &47.47(13)&0.01952(4)&\\
t$_{12}$ \cite{beuc2007absorption}&2065.14 &47.01&0.019     &2041.64 \\ \hline
\multicolumn{5}{c}{$X^2\Sigma_g^+$}\\
t$_{12}$& 6223.81 &46.37&0.017&6200.63\\
t$_{11}$& 6121.45 &45.42&0.017&6098.74\\ 
e$_5$   &   &  -  &  -   &6158.2$\pm 0.6$ \\ \hline \hline
\end{tabular}
\caption{Well depth $D_e$, vibrational constant $\omega_e$,  rotational constant $B_e$, and  dissociation energy of the lowest vibrational level $D_0=D_e-\omega_e/2$, of the $X$ and $B$ PECs of Rb$_2$, and of the $X^+$ PEC of Rb$_2^+$, used in the present analysis. Experimental values are reported \cite{amiot1990laser,strauss2010hyperfine}, as well as our calculations with the present FCI approach (t$_{12}$), and with the MRCI approach using the MOLPRO package \cite{werner2020} (t$_{11}$, see Section \ref{sec.discussion}), and our measured value e$_5$ (see Section \ref{sec:ionization}).}
\label{tab:Rb2pmin}
\end{table}

\section{Experimental set-up}
\label{sec:experiment}

In the experiment, a beam of Rb$_2$ is produced by the supersonic expansion of Rb vapor through a nozzle whose diameter is $300 ~\mu m$. The oven is at $625^0$ C, which corresponds to a rubidium vapor pressure of about 500 Torr. The nozzle is at a higher temperature of $675^0$ C to avoid clogging. This system is kept inside a source vacuum chamber at $ \sim \SI{1E-5}{Torr}$. The supersonic jet is collimated by a 5~mm diameter skimmer located 8.15~cm from the nozzle. The molecular beam intersects at the right angle with the excitation lasers 32.5~cm from the skimmer inside the detection chamber (pressure $ \sim \SI{1E-6}{Torr}$). This intersection region is located at the center of a metallic plate and a metallic ring, which guarantees a well-defined electric field strength ($F$). The expected inhomogeneity of the electric field strength in the axial direction is $\approx 11\%$ and in the radial direction is $\approx 3\%$, as determined by finite element analysis using the COMSOL Multiphysics software \cite{comsol_acdc}. More details are provided in the Supplemental Material (SM) \cite{supp}.

The ground-state molecules are detected by resonantly enhanced 2-photon ionization (RE2PI) using two different wavelengths (Fig. \ref{Fig_PECs}). The first photon excites the transition $X^1\Sigma_g^+ (v_X = 0)$ $\rightarrow$ $B^1\Pi_u (v_B = 2)$, which is chosen for better isotopic separation. The second photon is scanned in a wavenumber range around the expected minimum of the $X^2\Sigma_g^+$ PEC.

To perform such an excitation, a delay generator (Stanford Research, model DG535) triggers the Nd:YAG laser, whose second harmonic is used to pump simultaneously two pulsed dye lasers; one is a NarrowScanK from Radiant dyes (red laser ph1) and another Jaguar from Continuum (yellow laser ph2). Both lasers produce pulses of 5~ns duration at 20~Hz repetition rate. The first photon wavenumber is set at 14753~cm$^{-1}$, with a line width of 0.07~cm$^{-1}$. The second photon is provided by the yellow laser, which is scanned in the 16720~cm$^{-1}$-16755~cm$^{-1}$ wavenumber range, with a line width of 0.15~cm$^{-1}$. Due to the width of the lasers, the rotational structure of the molecule is not resolved. However, the red laser ph1 linewidth limits the excitation of a few rotational states. Wavelength calibration is performed using a wavemeter (High Finesse, model WS5), with an accuracy of 0.1~cm$^{-1}$. The optical excitation occurs at a zero electric field. The same delay generator triggers a high voltage (HV) pulse generator (DEI, model PVX-4140), which is fed by an HV power supply (Stanford Research, model PS350). A positive voltage pulse, whose rise time is 20~ns, is applied in one metal grid 1.13~$\mu s$ after the optical excitation while the other grid is grounded. 

As illustrated in Fig.~\ref{Fig_PECs}, the yellow photon intends to directly photoionize Rb$_2$ molecules excited in the $B$ state, or to populate levels of molecular Rydberg states which could autoionize. We extract the produced Rb$_2^+$ ions with a pulsed electric field that lowers the ionization threshold of the Rydberg molecular states \cite{Gallagher_1994,Robicheaux2000Ionization}. This pulse field ionization technique (PFI) relies on an electric field that gradually increases in strength over time and ultimately detaches the electron from the atom/molecule, forming an ion. It has recently been used to study Mg$_2^+$ and MgKr$^+$ molecular ions \cite{GENEVRIEZ2022111591,Kreis17092023}. The HV pulse accelerates the ions to a double electrostatic field time-of-flight lens system for detection by a channeltron.

\begin{figure*}[t]
\centering
\includegraphics[width=0.90\textwidth]{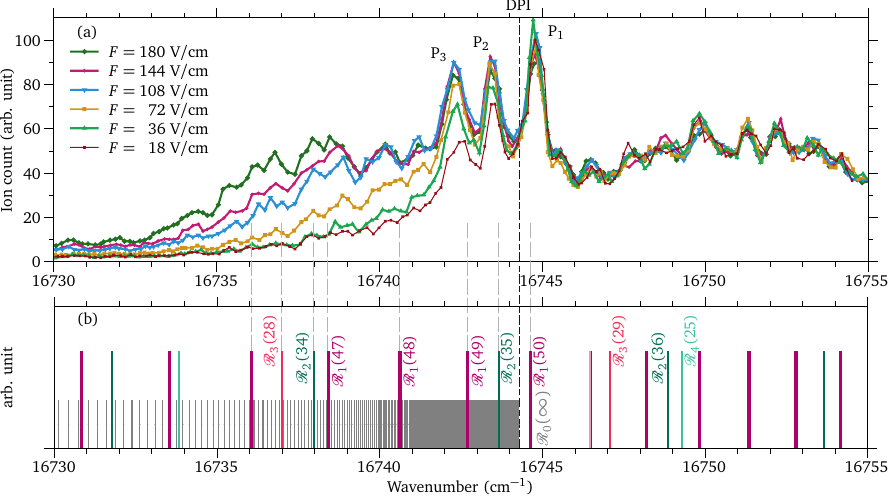}
\caption{\label{Fig_Spectrum} (Color online) (a) Normalized Rb$_2^+$ ion count as a function of the yellow photon wavenumber for several electric field intensities $F$. Three prominent peaks are identified as P$_1$, P$_2$, and P$_3$. The onset of direct photo-ionization (DPI) is located at 16744.3~cm$^{-1}$. (b) Transition lines (with arbitrary height) from the present model for molecular Rydberg series $n$ converging toward the vibrational levels $v = 0,1,2, ...$ of the $^{85}$Rb$_2^+$  ground state; they are labeled as $\mathscr{R}_v(n)$. A tentative assignment is indicated with vertical dashed lines.}
\end{figure*}

\section{Recorded spectra and analysis}
\label{sec:spectra}

\begin{figure}[h]
\centering
\includegraphics[width=0.475\textwidth]{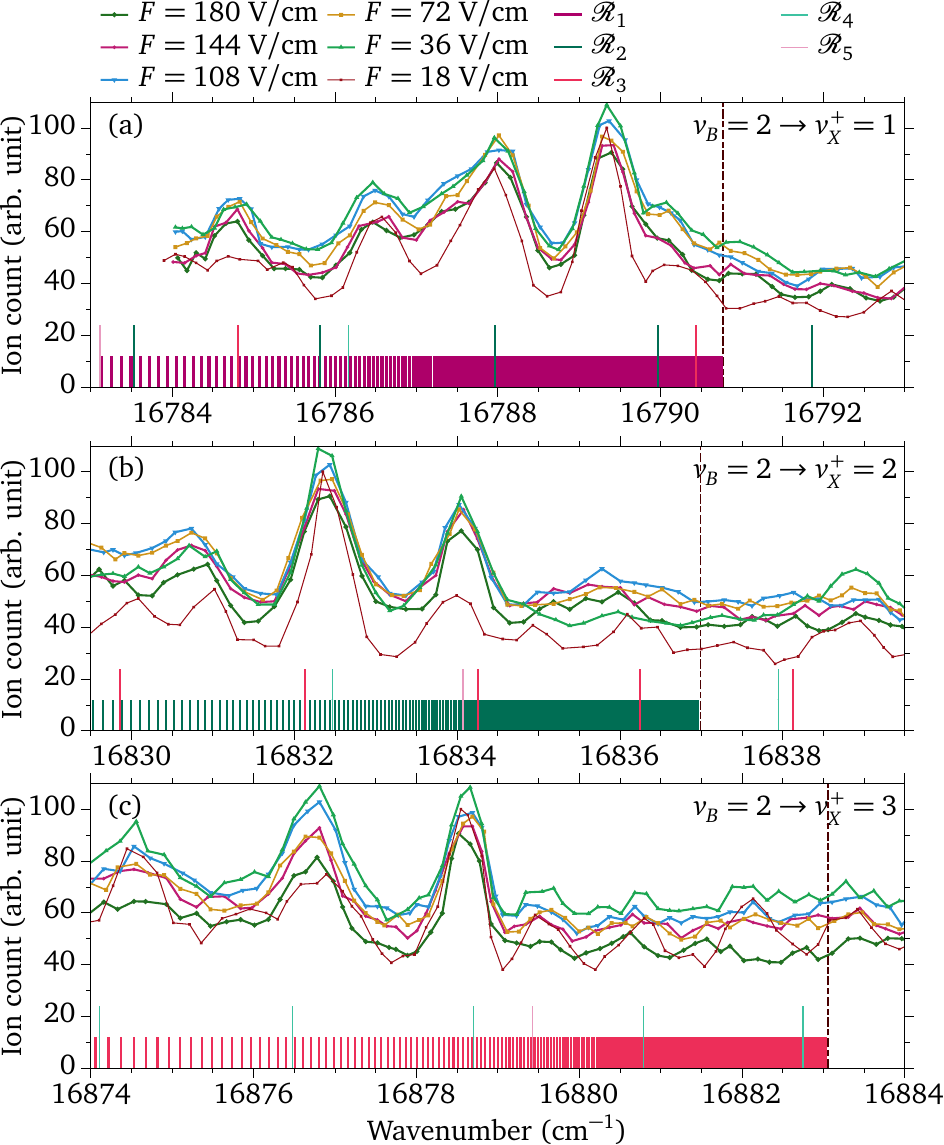}
\caption{ (Color online)  Normalized Rb$_2^+$ ion count as a function of the yellow photon wavenumber for several electric field strengths $F$ near the expected $v=1$ (a), $v=2$ (a), $v=3$ (c) levels of Rb$_2^+$. Modeled Rydberg series $\mathscr{R}_v$ ($v=1$ to 5) are also reported with arbitrary height.}
\label{Figexep123}
\end{figure}

The recorded Rb$_2^+$ spectra (Fig. \ref{Fig_Spectrum}(a)) is plotted as a function of the yellow photon wavenumber for several electric fields ($F =$ 180, 144, 108, 72, 36, and 18 V/cm). The signal is normalized to the total ion counts recorded in the range 16746.0~cm$^{-1}$-- 16754.5~cm$^{-1}$: all spectra clearly overlap in this region, independently of the field strength. Below this range, we see an enhancement of the signal background with increasing electric field strength, confirming the influence of the PFI on field-free Rydberg states. The separation between these two parts signs the onset of the direct photoionization (DPI) signal that occurs when the yellow photon energy is large enough to reach the ground vibrational level $v=0$ of the $X^+$ state. The precise determination of the energy position of this DPI requires a deeper analysis of the spectra (Section \ref{sec:ionization}). Spectra extended toward the energy of the $v=1,2,3$ thresholds with the same normalization as that of Fig. \ref{Fig_Spectrum}(a) are provided in Fig. \ref{Figexep123}.

Three prominent peaks, P$_1$, P$_2$, and P$_3$, are clearly visible in Fig. \ref{Fig_Spectrum} (a), and some others in the extended parts of the spectra provided in the SM. Such peaks are not expected for DPI in the electronic energy continuum. They are evenly spaced in energy, by much larger spacings than the rotational structure of the molecule. Therefore, these peaks should be associated with the Rb$_2$ Rydberg vibrational levels. We fitted them with three Gaussian functions for each electric field strength, producing the same series of mean wavenumbers 16744.8 cm$^{-1}$ (P$_1$),  16743.5 cm$^{-1}$ (P$_2$), and 16742.4 cm$^{-1}$ (P$_3$). To assess this hypothesis, we set up a simple quantum defect model, defining the energy of the Rydberg level $n$ belonging to the series $\mathscr{R}_v$ that converges towards the vibrational level $v$ of Rb$_2^+(X^+)$ as
\begin{equation}
E_v(n) = E_v -  \frac{R_{\infty} ( 1 + \frac{m_e}{m_d})^{-1}}{(n^*)^2},
\label{Eq:Ryd}%
\end{equation}
where $R_{\infty}$ is the Rydberg constant, corrected to account for the finite mass $m_d$ of $^{85}$Rb$_2^+$ ($m_e$ being the electron mass), and $n^\star = n - \delta_v$ \cite{li2003millimeter} involving the quantum defect $\delta_v$ of the molecular Rydberg series $\mathscr{R}_v$ depending on $v$. 

In Fig. \ref{Fig_Spectrum}(b), we show our calculated spectrum with vertical lines of arbitrary height. The ionization threshold of the $\mathscr{R}_0$ series (small gray lines), namely the position of the Rb$_2^+(X^+)$ $v=0$ level, is fixed at 16744.3~cm$^{-1}$ (see Section \ref{sec:ionization}). The corresponding Rydberg levels labeled $\mathscr{R}_0(n)$ can only be ionized by PFI. Their density is so high in this region that they contribute to the background ion signal. Then in the model, the next series $\mathscr{R}_v$ converge towards the calculated position of the consecutive vibrational levels of the $X^+$ PEC. For instance, $E_1=E_0+46.37$~cm$^{-1}$, $E_2=E_1+46.22$~cm$^{-1}$, ..., consistently with the spectroscopic constants reported in Table \ref{tab:Rb2pmin}.

The fitted positions of the levels $\mathscr{R}_v(n)$ belonging to the series $\mathscr{R}_v$ are represented by colored vertical bars for $v>0$. Such levels lie below the $v=0$ threshold so that they are ionized by PFI (\textit{i.e.} P$_2$ assigned to $\mathscr{R}_2(35)$ and P$_3$ to $\mathscr{R}_1(49)$), or above $v=0$ and then autoionize (\textit{i.e.} P$_1$ assigned to $\mathscr{R}_1(50)$). The quantum defect $\delta_1$ of the series $\mathscr{R}_1$ is optimized to reproduce lines recorded in the range of Fig. \ref{Fig_Spectrum}(a). 
The quantum defects $\delta_v$ ($v=2,3,4$) are fitted over the energy ranges shown in Fig. \ref{Fig_Spectrum}(a) \& Fig. \ref{Figexep123}(a); Fig. \ref{Figexep123}(a) \& Fig. \ref{Figexep123}(b); Fig. \ref{Figexep123}(b) \& Fig. \ref{Figexep123}(c), respectively.
%
%
Finally, $\delta_5$ is optimized on the single spectrum of Fig. \ref{Figexep123}(c). The final values are $\delta_1=1.17$, $\delta_2=0.69$, $\delta_3=0.58$, $\delta_4=0.28$, $\delta_5=0.07$. Despite the simplicity of our model, we can unambiguously assign the observed lines to the manifestation of molecular Rydberg levels. We note a strong variation of $\delta_v$ with $v$. This is surprising since only low values of $v$ are considered, namely, levels close to the bottom of the $X^+$ PEC for which the quantum defect is not expected to strongly vary with such a limited range of internuclear distance. This pattern may be due to the presence of a doubly-excited autoionizing state that would perturb the Rydberg series, as seen, for example, in Na$_2$ \cite{dulieu1991}.

\section{Ionization energy of R$\lowercase{\textrm{b}}$$_2$ }
\label{sec:ionization}

\begin{figure}[t]
\centering
\includegraphics[width=0.475\textwidth]{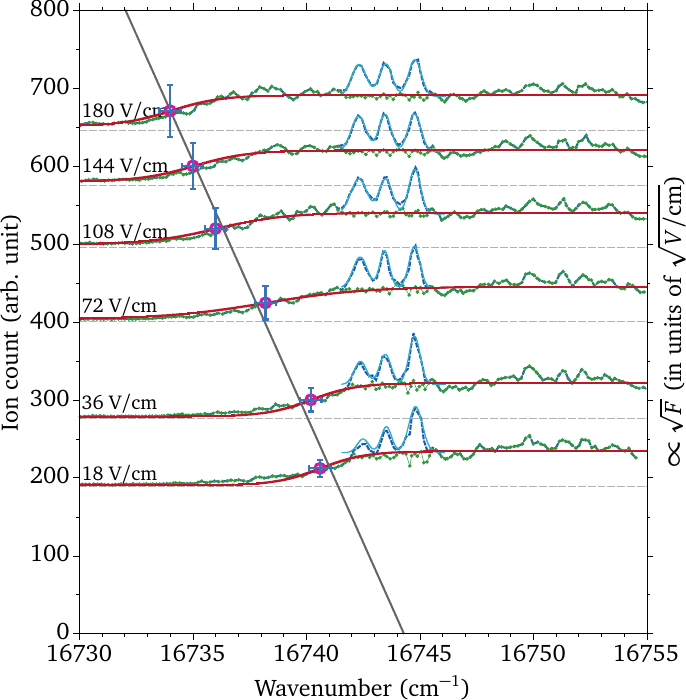}
\caption{\label{fig:Fig_analysis} The normalized Rb$_2^+$ spectra as a function of wavenumber $\tilde{\nu}$ shown in Fig. \ref{Fig_Spectrum}(a) (green lines) for several electric field strengths $F$, after removal of the contributions from the peaks P$_1$, P$_2$, and P$_3$ (blue lines), and fitted with the logistic function $I(\tilde{\nu})$ defined in the text (red lines). The spectra are vertically shifted proportionally to $\sqrt{F}$. The half-rise of the signal is marked with open circle, with horizontal and vertical error bars. The linear extrapolation to $F=0$ yields the Rb$_2$ ionization energy $E_i=16744.3 \pm 0.5$~cm$^{-1}$.} 
\end{figure}

The second photon wavenumber that corresponds to the field-free ionization energy $E_i$ of the $v=0$ level of the Rb$_2$ ground state is reported in Fig. \ref{Fig_Spectrum}(a) as a vertical dashed line labeled DPI. The value of $E_i$ is deduced from the analysis of the behavior of the onset of the ionization signal as a function of the applied electric field strength. We follow the approach of \cite{Genevriez2019} applied to the PFI of Mg$^+$. In the present molecular case, we first removed the contribution of the peaks P$_1$, P$_2$ and P$_3$ from the spectrum (blue lines in Fig. \ref{fig:Fig_analysis}, using their Gaussian modeling presented above. The intensity $I({\tilde\nu})$ of remaining PFI spectrum (green lines in Fig. \ref{fig:Fig_analysis}) is then fitted for each electric field strength $F$ as a function of the wavenumber $\tilde\nu$ with the logistic function (red lines in Fig. \ref{fig:Fig_analysis})
\begin{equation}
I({\tilde\nu})=I_{\textrm{min}}+\frac{I_{\textrm{max}}}{1+exp(-\beta(\tilde\nu-\tilde\nu_0))},
\label{eq:logistic}
\end{equation}
where, for a given intensity $F$, $I_{\textrm{min}}$ is the initial background, $I_{\textrm{max}}$ is the maximal intensity of the spectrum, $\tilde\nu_0$ is the wavenumber of the half-rise of the signal (open circles in Fig. \ref{fig:Fig_analysis}), $\beta$ the sharpness of the step function. The values of the couple ($I_{\textrm{min}}$, $I_{\textrm{max}}$) are (1,44), (2,43), (3,42), (4,41), (5,40), (5,40), for $F = 18, 36, 72, 108, 144, 180$~V/cm, respectively. The onset of ionization is shifted downward by the electric field strength by an amount proportional to $\sqrt{F}$ \cite{Gallagher_1994}. In Fig. \ref{fig:Fig_analysis}, the spectra are ordered with increasing values of $F$, and are shifted along the vertical axis by an offset proportional to $\sqrt{F}$. The vertical error bars at $\tilde{\nu}_0(\sqrt{F})$ mirror the relative uncertainty of $F$ (estimated at $\pm 10\%$). The horizontal error bars conservatively cover $\pm 5\sigma$, where $\sigma = 0.09$~cm$^{-1}$ is derived from the sum of the half-line widths of the two lasers. The linear regression of $\tilde{\nu}_0(\sqrt{F})$ extrapolated to $F=0$ yields the wavenumber $\tilde{\nu}_0(0)=16744.3 \pm 0.5$~cm$^{-1}$ of the yellow laser at which the $v=0$ level of the $X^+$ state is reached at a zero field. By adding the wavenumber of the red laser to $\tilde{\nu}_0(0)$ ($14753.0 \pm 0.03$~cm$^{-1}$), we obtain the ionization energy of $^{85}$Rb$_2$ $E_i = 31497.3\pm 0.6$ cm$^{-1}$. We note that the coefficient of linear regression is 0.767, compared to 6.12 in \cite{Genevriez2019}. This value strongly depends on the time derivative of the pulsed field \cite{Robicheaux2000Ionization}.

We extract the dissociation energy $D_0(X^+)$ from the $X^+$ PEC of $^{85}$Rb$_2^+,X^2\Sigma_g^+ (v=0)$ according to 
\begin{equation}
D_0(X^+)=E^+(\textrm{Rb})+D_0(X)-E_i,
\label{eq:D0}
\end{equation}
where $E^+(\textrm{Rb})=33690.81$~cm$^{-1}$ is the measured $^{85}$Rb ionization threshold \cite{johansson1961spectra} which fixes the difference between the asymptotic limit Rb($5s$)$+$Rb($5s$) of Rb$_2$ ground state and the asymptotic limit  Rb$^+$+Rb($5s$) of  Rb$_2^+$ ground state, and  $D_0(X) = 3964.718$~cm$^{-1}$ is the  measured dissociation energy of the $^{85}$Rb$_2$ $X^1\Sigma_g^+$ ground state (Table \ref{tab:Rb2pmin}). We find $D_0(X^+)= 6158.2\pm 0.6$~cm$^{-1}$.

\section{Discussion and conclusion}
\label{sec.discussion}

\begin{figure}[t]
\centering
\includegraphics[width=0.465\textwidth]{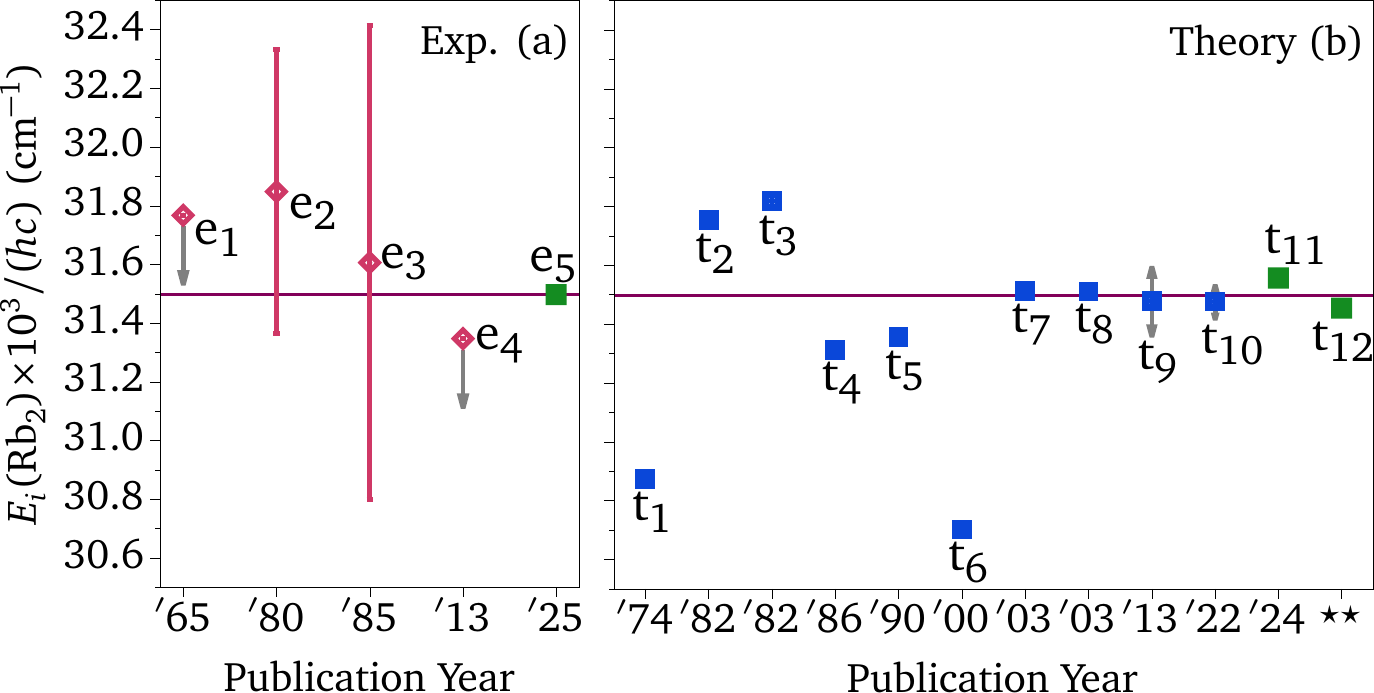}
\caption{\label{Fig-expth-ref} (a) Experimentally measured ionization energies of Rb$_2$$X^1\Sigma_g^+$. The present measurement, e$_5$, is shown by a square. A horizontal line at the corresponding energy, $E_i = 31497.3$ cm$^{-1}$, highlights the comparison with previous results. Refs: e$_1$\cite{lee1965photosensitized}, e$_2$\cite{klucharev1980ionisation}, e$_3$\cite{wagner1985formation}, e$_4$\cite{bellos2013upper} (b) Ionization energies of Rb$_2$ computed using different methods. Refs:  t$_1$\cite{bellomonte1974alkali}, t$_2$\cite{von1982pseudopotential}, t$_3$\cite{jeung1982inclusion}, t$_4$\cite{silberbach1986ground}, t$_5$\cite{krauss1990effective}, t$_6$\cite{patil2000simple}, t$_7$\cite{jraij2003theoretical}, t$_8$\cite{aymar2003model}, t$_9$\cite{bellos2013upper}, t$_{10}$\cite{schnabel2022high}, t$_{11}$\cite{pandey2024ultracold}, t$_{12}$\cite{dasilva2024}  }
\end{figure}

In Fig. \ref{Fig-expth-ref} we collected
representative published experimental values with their error bars (e$_i$, panel (a)) and theoretical values (t$_j$, panel (b)), for the ionization energy $E_i$ of the $^{85}$Rb$_2$ ground state. The measured value of the present work appears as e$_5$, and the calculated ones as t$_{11}$ and t$_{12}$. It should be noted first that the t$_j$ values are not directly computed as ionization energies. They are based on the calculation of the dissociation energy $D_0(X^+)$ of the $v=0$ level of the $X^+$ PEC (see Table \ref{tab:Rb2pmin}), which is then converted into the ionization energy $E_i=E(5s)+D_0(X)-D_0(X^+)$, where $E(5s)$ is the $^{85}$Rb ionization energy, and $D_0(X)$ the dissociation energy of the $v=0$ level of the $X$ PEC (Table \ref{tab:Rb2pmin}). Most of the reported theoretical values of $D_0(X^+)$ result from the modeling of Rb$_2^+$ as a one-valence-electron system in the field of various forms of ECP and various basis sets (t$_1$\cite{bellomonte1974alkali}, t$_2$\cite{von1982pseudopotential}, t$_3$\cite{jeung1982inclusion}, t$_4$\cite{silberbach1986ground}, t$_5$\cite{krauss1990effective}, t$_6$\cite{patil2000simple}, t$_7$\cite{jraij2003theoretical}, t$_8$\cite{aymar2003model}, t$_9$\cite{bellos2013upper}). A noticeable difference comes with the t$_{10}$ value computed with the coupled-cluster method \cite{schnabel2022high}. All the most recent determinations, t$_7$ to t$_{12}$, thus including ours, with different approaches, nicely converge around the same value, in good agreement with our measured e$_5$ value.

While our e$_5$ value for the ionization energy is consistent with the experimental values e$_1$, e$_2$, and e$_3$ with large error bars, it appears larger by $149.3$~cm$^{-1}$ than the upper bound obtained in \cite{bellos2013upper}. In the latter experiment, the excitation is different, as is the recorded spectrum. A pair of cold Rb atoms is associated in a cold Rb$_2$ molecule created at a weakly-bound vibrational level of the $a^3\Sigma_u^+$ state correlated to the Rb($5s$)+ Rb($5s$) dissociation limit. These molecules are excited by uv light with wavelength around 365~nm into vibrational levels lying close to the Rb($5s$)+ Rb($7p$), which has an energy presumably higher than the level $v=0$ of the $X^+$ state. In contrast to our experiment, the recorded Rb$_2^+$ spectrum starts with sharp lines, indicating the onset of ionization as soon as a vibrational level has enough energy to autoionize. This line appears in an energy range for which, in our experiment, even when an electric field strength of about 360~V/cm is applied, no ions are detected (see the corresponding spectrum in the SM). We infer from \cite{CarolloPhD} that an electric field strength at least 300~V/cm was present in \cite{bellos2013upper} to explain the appearance of molecular ions that would have been induced by field ionization. The value of the ionization energy of Rb$_2$ obtained in the present work completes the knowledge of this quantity for all alkali dimers \cite{stwalley1993b}. The present results open the way to further study on Rb$_2$ molecular Rydberg state, for instance using ZEKE spectroscopy \cite{GENEVRIEZ2022111591}. Such studies could provide perspectives for research on ultra-long-range Rydberg systems \cite{greene2000,farooqi2003,booth2015,shaffer2018,fey2020}, for example, to create a novel triatomic system \cite{londono2025} involving a Rydberg Rb$_2$ molecule and a Rb atom. 

\begin{acknowledgments}
Enlightening discussions with Fr\'ed\'eric Merkt and Matthieu Génévriez, Michael Bellos and Phil Gould are gratefully acknowledged. This work is supported by Grants 2018/06835-0, 2022/16904-5, 2023/06732-5, and 2021/04107-0 from São Paulo Research Foundation (FAPESP), FA9550-23-1-0666 from the US Air Force Office of Scientific Research, 305257/2022-6 from CNPq, and ANR-21-CE30-0060-01 (COCOTRAMOS project) from the Agence Nationale de la Recherche. 
\end{acknowledgments}

\bibliography{References}

\end{document}